\documentclass[a4paper]{jpconf}
\usepackage{graphicx}

\usepackage{amssymb}

\newcommand{\apjl}{\emph{Astrophys. J. Lett.} }
\newcommand{\apj}{\emph{Astrophys. J.} }

\newcommand{\app}{\emph{Astropart. Phys. }}

\newcommand{\prd}{\emph{Physics Rev. D } }

\begin{document}
\title{Latest results from the ARGO-YBJ experiment}

\author{Di Sciascio Giuseppe on behalf of the ARGO-YBJ Collaboration}

\address{INFN - Sezione Roma Tor Vergata, Roma, Italy}
\ead{giuseppe.disciascio@roma2.infn.it}

\begin{abstract}
The ARGO-YBJ experiment has been in stable data taking for 5 years at the YangBaJing Cosmic Ray Observatory (Tibet, P.R. China, 4300 m a.s.l., 606 g/cm$^2$). With a duty-cycle greater than 86\% the detector collected about 5$\times $10$^{11}$ events in a wide energy range, from few hundreds GeV up to about 10 PeV.
A number of open problems in cosmic ray physics has been faced  exploiting different analyses.
In this paper we summarize the latest results in cosmic ray physics and in gamma-ray astronomy.
\end{abstract}

\section{Introduction}

Aiming to face the open problems in Galactic cosmic ray (hereafter CR) physics through a combined study of photon- and charged particle-induced extensive air showers (hereafter EAS) with the same detector, ARGO-YBJ has been in stable data taking for more than 5 years at the YangBaJing Cosmic Ray Observatory (Tibet, P.R. China, 4300m a.s.l., 606 g/cm$^2$, 90.5$^{\circ}$ East, 30.1$^{\circ}$ North). 
The detector acted simultaneously as a wide aperture ($\sim$sr), continuosly-operated $\gamma$-ray telescope at sub-TeV -- TeV photon energies and as a high resolution CR detector in the broad energy range between few TeV and 10 PeV.

The combination of the high elevation of the site, the full coverage of the central carpet and the high granularity of the readout provides very important advantages in various aspects of CR physics. 
In fact, in addition to the decrease of the threshold energy down to few hundreds GeV, the location of the esperiment above 4000 m a.s.l. ensures that shower fluctuations are small (we are working in the shower maximum region) and that all nuclei produce showers with nearly the same electromagnetic size.
The low energy threshold is crucial for the overposition with direct measurements carried out by balloon/satellite-born detectors allowing one to cross-calibrate the different energy scales. The nearly independence of the size on the mass of the primary determinates the size -- energy relation to be better defined, allowing the study of the elemental composition around the knee in a very efficient and reliable way.

In this paper the latest results obtained by ARGO-YBJ in CR physics and in gamma-ray astronomy will be briefly summarized.

\section {The ARGO-YBJ experiment}
\label{}

ARGO-YBJ is a full coverage air shower detector constituted by a central carpet $\sim$74$\times$78 m$^2$, made of a single layer of resistive plate chambers (RPCs) with $\sim$93$\%$ of active area, enclosed by a guard ring partially instrumented ($\sim$20$\%$) up to $\sim$100$\times$110 m$^2$. The apparatus has a modular structure, the basic data acquisition element being a cluster (5.7$\times$7.6 m$^2$), made of 12 RPCs (2.85$\times$1.23 m$^2$ each). Each chamber is read by 80 external strips of 6.75$\times$61.80 cm$^2$ (the spatial pixels), logically organized in 10 independent pads of 55.6$\times$61.8 cm$^2$ which represent the time pixels of the detector \cite{aielli06}. 
The readout of 18,360 pads and 146,880 strips is the experimental output of the detector. 
In addition, in order to extend the dynamical range up to PeV energies, each chamber is equipped with two large size pads (139$\times$123 cm$^2$) to collect the total charge developed by the particles hitting the detector \cite{bigpad}.
The RPCs are operated in streamer mode by using a gas mixture (Ar 15\%, Isobutane 10\%, TetraFluoroEthane 75\%) for high altitude operation \cite{bacci00}. The high voltage settled at 7.2 kV ensures an overall efficiency of about 96\% \cite{aielli09a}.
The central carpet contains 130 clusters and the full detector is composed of 153 clusters for a total active surface of $\sim$6,700 m$^2$. The total instrumented area is $\sim$11,000 m$^2$.
For each event the location and timing of every detected particle is recorded, allowing the reconstruction of the lateral distribution and the arrival direction. The trigger is based on a time correlation among the pad signals depending on their relative distance. In this way, all the shower events giving a number of fired pads N$_{pad}\ge$ N$_{trig}$ in the central carpet in a time window of 420 ns generate the trigger. 
The whole system has been in stable data taking from November 2007 to January 2013, with the trigger condition N$_{trig}$ = 20 and a duty cycle $\geq$86\%. The trigger rate is $\sim$3.5 kHz with a dead time of 4$\%$.

Because of the small pixel size, the detector is able to record events with a particle density exceeding 0.003 particles m$^{-2}$, keeping good linearity up to a core density of about 15 particles m$^{-2}$.
This high granularity allows a complete and detailed three-dimensional reconstruction of the front of air showers at an energy threshold of a few hundreds GeV. Showers induced by high energy primaries ($>$ 100 TeV) are also imaged by the charge readout of the large size pads which allows to study the structure of the shower core region up to particle densities of $\sim$10$^{4}$/m$^2$ \cite{bigpad}.

Details on the analysis procedure (e.g., reconstruction algorithms, data selection, background evaluation, systematic errors) are discussed in \cite{aielli10,bartoli11a,bartoli11b}.
The performance of the detector (angular resolution, pointing accuracy, energy scale calibration) and the operation stability are continuously monitored by observing the Moon shadow, i.e., the deficit of CRs detected in its direction \cite{bartoli11a,bartoli12a}. 
The measured angular resolution is better than 0.5$^{\circ}$ for CR-induced showers with energy E $>$ 5 TeV and the overall absolute pointing accuracy is $\sim$0.1$^{\circ}$.  According to MonteCarlo (MC) simulations the angular resolution for $\gamma$-induced events results smaller by 30\% - 40\%. The absolute rigidity scale uncertainty of ARGO-YBJ is estimated at 10\% level in the range 1 - 30 TeV/Z \cite{bartoli11a,bartoli12a}.

\section{Cosmic Ray Anisotropy}

The CR arrival direction distribution and its anisotropy has been a long-standing problem ever since the 1930s. The study of the anisotropy is complementary to the study of their energy spectrum and elemental composition to understand CR origin and propagation. It is also a powerful tool to probe the structure of the magnetic fields through which CRs travel.

The anisotropy in the CR arrival direction distribution hs been observed by different experiments with increasing sensitivity and details at different angular scales.
Current experimental results show that the main features of the anisotropy are uniform in the energy range (10$^{11}$ - 10$^{14}$ eV), both with respect to amplitude (10$^{-4}$ - 10$^{-3}$) and phase ((0 - 4) hr).
The existence of two distinct broad regions, one showing an excess of CRs (called ``tail-in''), distributed around 40$^{\circ}$ to 90$^{\circ}$ in R.A., the other a deficit (the ``loss cone''), distributed around 150$^{\circ}$ to 240$^{\circ}$ in R.A., has been clearly observed (for a review see, for example, \cite{disciascio13}).
In the last years different experiments reported evidence of the existence of a medium angular scale anisotropy in the both hemispheres \cite{disciascio13}.

\begin{figure}
\begin{minipage}[ht]{.47\linewidth}
  \centerline{\includegraphics[width=\textwidth]{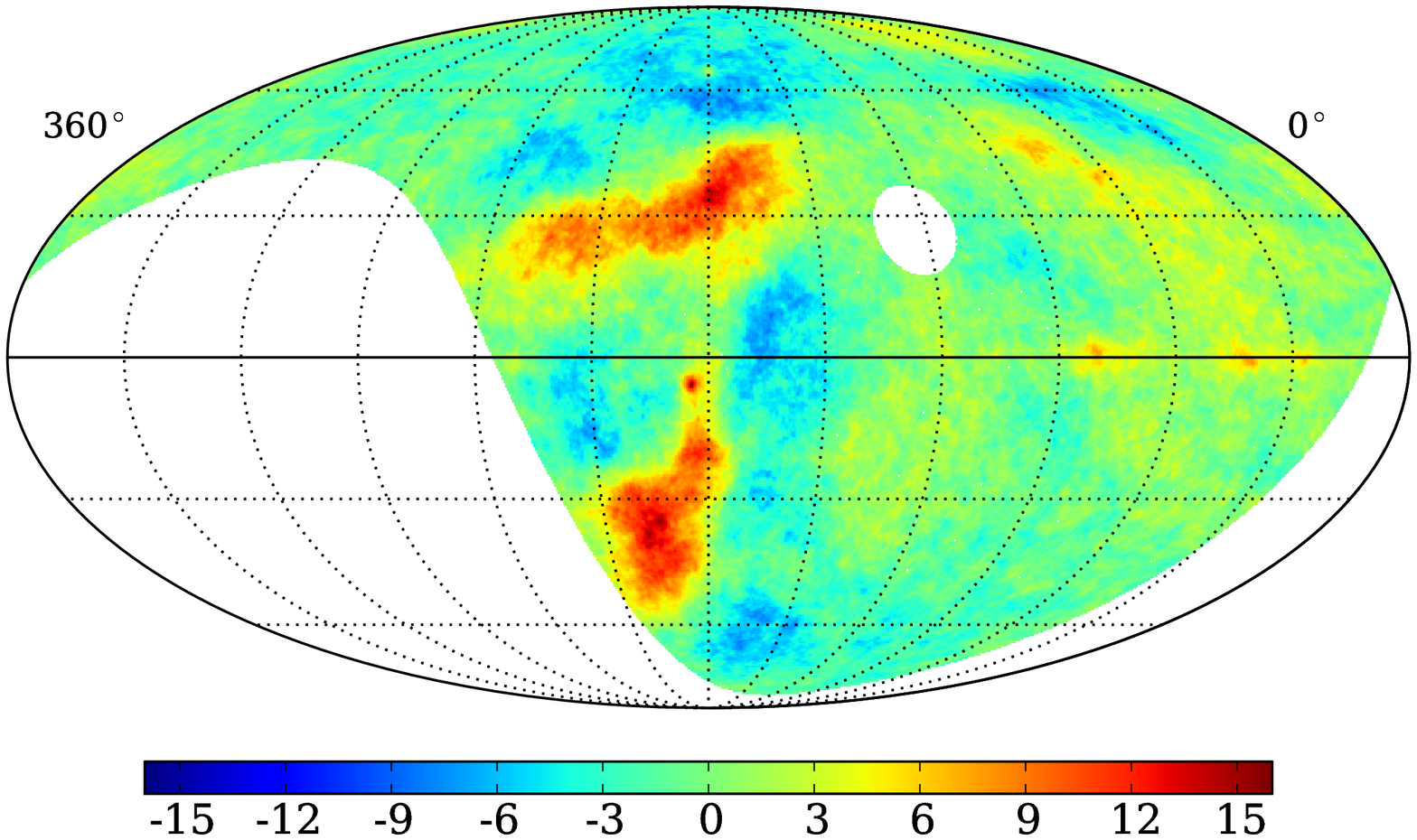}}
    \caption{ARGO-YBJ sky-map in galactic coordinates. The color scale gives the statistical significance of the observation in standard deviations. The map center points towards the galactic Anti-Center.}
\label{fig:fig10}
\end{minipage}\hfill
\begin{minipage}[ht]{.47\linewidth}
  \centerline{\includegraphics[width=\textwidth]{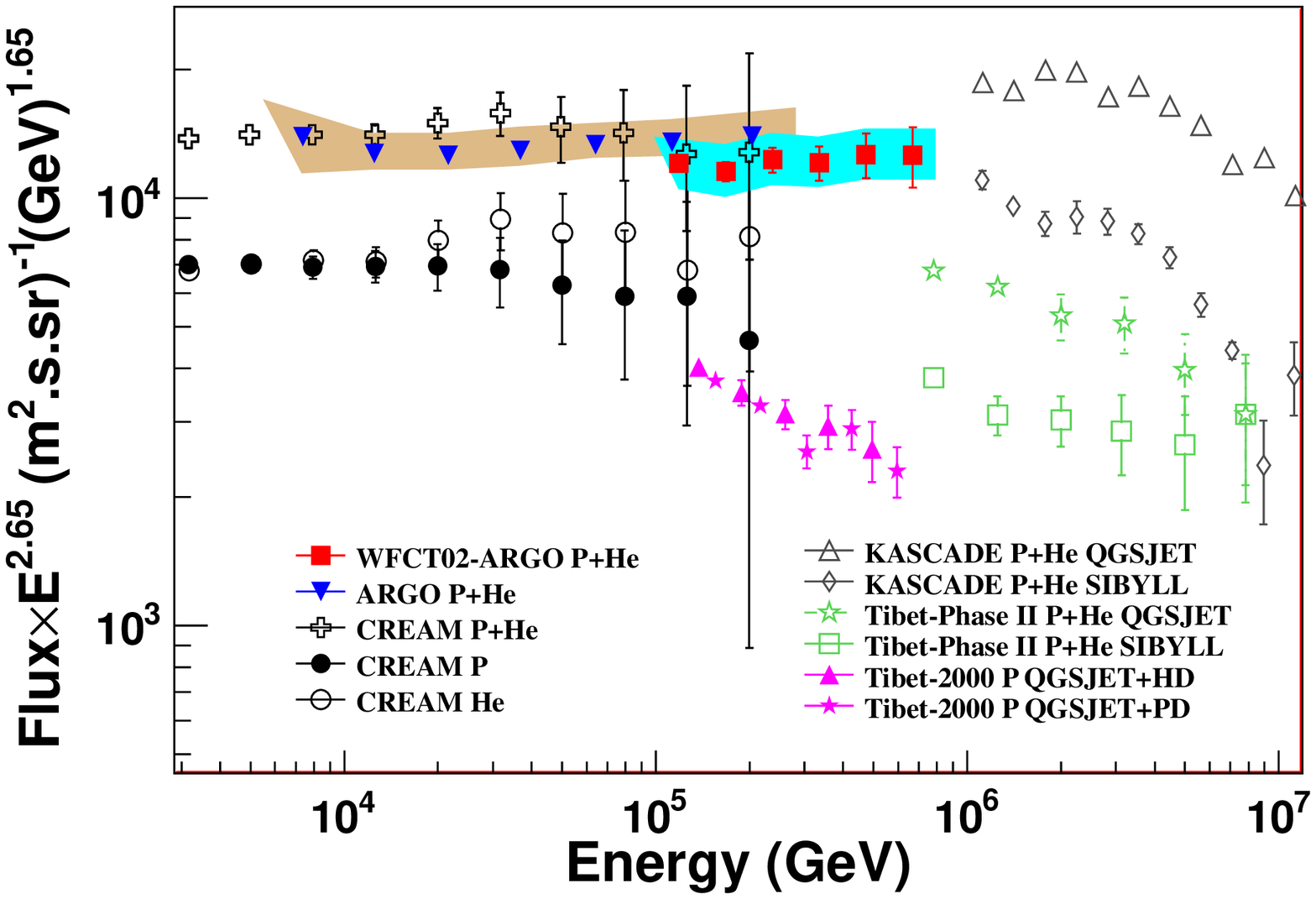} }
    \caption{Light component (p+He) energy spectrum of primary CRs measured by ARGO-YBJ compared with other experimental results. The results obtained by the ARGO-YBJ/WFCTA system are shown by the filled red squares \cite{argo-wfcta}.}
 \label{fig:light_spectrum}  
        \end{minipage}\hfill
\end{figure}

In Fig. \ref{fig:fig10} the ARGO-YBJ sky map of medium angular scale (order of 10$^{\circ}$) anisotropy in galactic coordinates as obtained with 4.5 years data is shown (for details see \cite{iuppa13,bartoli13b}). The map center points towards the galactic Anti-Center. 
The zenith angle cut ($\theta\leq$ 50$^{\circ}$) selects the declination region $\delta\sim$ -20$^{\circ}\div$ 80$^{\circ}$.
In this zenith angle bin the angular resolution is nearly constant and the energy of selected events is in the TeV range.
According to the simulation, the median energy of the isotropic CR proton flux is E$_p^{50}\approx$1.8 TeV (mode energy $\approx$0.7 TeV).
No gamma/hadron discrimination algorithms have been applied to the data. Therefore, the sky map is filled with all CRs possibly including photons, without any discrimination.

In spite of the fact that the bulk of Supernova Remnants (SNRs), pulsars and other potential CR sources are in the Inner Galaxy surrounding the Galactic Centre, the excess of CR is observed in the opposite, Anti-Centre direction. As stressed in \cite{erlykin13}, the fact that the observed excesses are in the Northern and in the Southern Galactic hemisphere, favors the conclusion that the CR at TeV energies originate in sources whose directions span a large range of Galactic latitudes.
The right side of the sky map shows different few-degree excesses not compatible with random fluctuations (the statistical significance is up to 7 s.d.). The observation of these structures is reported by ARGO-YBJ for the first time, as discussed in \cite{bartoli13b}.

From the theoretical viewpoint, the anisotropy of CRs is important as a direct trace of potential sources. Nonetheless, models failed to explain the whole set of observations at different angular scales \cite{disciascio13}.
Unlike predictions from diffusion models, that foresee only a dipole, the CR arrival distribution in sidereal time was never found to be purely dipolar, what translates in needing at least two harmonics to describe the data (the observation of anisotropies down to $\sim$10$^{\circ}$ implies multipoles of order 18). 
So far, no theory of CRs in the Galaxy exists which is able to explain the origin of these different anisotropies leaving the standard model of CRs and that of the local Galactic Magnetic Field unchanged at the same time.
The anisotropy problem is the most serious challenge to the standard model of the origin of galactic CRs from diffusive shock acceleration \cite{hillas05}.

%%%%%%%%%%%%%%%%%%%%%%%%%%%%%%%%%%%%%%%%
%%%%%%%%%%%%%%%%%%%%%%%%%%%%%%%%%%%%%%%%

\section{Measurement of the Cosmic Ray Energy spectrum}

There is a general consensus that Galactic CRs up to the all-particle knee originate in SNRs accelerated by the first order Fermi mechanism in shock waves. The theoretical modelling of this mechanism can reproduce, in principle, the measured spectra and composition of CRs. 
Recently AGILE and Fermi observed GeV photons from two young SNRs (W44 and IC443) showing the typical spectrum feature around 1 GeV (the so called ``$\pi^0$ bump'', due to the decay of $\pi^0$) related to hadronic interactions \cite{pizero-a,pizero-f}. 
This important measurement however does not demonstrate the capability of SNRs to accelerate CRs up to the knee and above. 

In the standard picture, mainly based on the results of the KASCADE esperiment, the knee is attributed to the steepening of the p and He spectra \cite{kascade}. 
However, a number of experiments reported evidence that the bending of the light component (p+He) is well below the PeV and the knee of the all-particle spectrum is due to heavier nuclei (see, for example, \cite{tibet,casamia,basje}. 

A large number of theoretical papers discussed the highest energies achievable in SNRs and the possibility that protons can be accelerated up to PeVs (for a recent review see \cite{blasi13} and references therein).
Therefore, the determination of the proton knee, as well as the measurement of the evolution of the heavier component across the knee, are the key components for understanding origin and acceleration mechanisms of Galactic CRs. 

A measurement of the CR primary energy spectrum (all-particle and light component) in the energy range few TeV -- 10 PeV is under way with the ARGO-YBJ experiment.
To cover this wide energy range different 'eyes' have been used:
\begin{itemize}
\item \emph{'digital readout'}, based on the strip multiplicity, i.e. the picture of the EAS provided by the strip/pad system, in the few TeV -- 200 TeV energy range \cite{bartoli12};
\item \emph{'analog readout'}, based on the particle density in the shower core region, in the 100 TeV -- 10 PeV range \cite{discia14};
\item \emph{'hybrid measurement'}, carried out by ARGO-YBJ and a wide field of view Cherenkov telescope, in the 100 TeV - PeV region \cite{argo-wfcta}.
\end{itemize}

In the following we summarize results obtained below 100 TeV with the digital readout and describe the hybrid measurement up to  the PeV range. Preliminary results obtained with the ARGO-YBJ analog data are described in \cite{discia14}.

\subsection{Light component (p+He) spectrum of Cosmic Rays}

As described in \cite{bartoli12}, requiring quasi-vertical showers ($\theta$ $<$ 30$^{\circ}$) and applying a selection criterion based on the particle density, a sample of events mainly induced by p and He nuclei, with shower core inside a fiducial area (with radius $\sim$28 m), has been selected. The contamination by heavier nuclei is found negligible. An unfolding technique based on the Bayesian approach has been applied to the strip multiplicity distribution in order to obtain the differential energy spectrum of the light component. 

The spectrum measured by ARGO-YBJ is compared with other experimental results in Fig. \ref{fig:light_spectrum} (blue inverted triangles).  
The ARGO-YBJ data agree remarkably well with the values obtained by adding up the p and He fluxes measured by CREAM both concerning the total intensities and the spectral index \cite{cream11}. The value of the spectral index of the power-law fit to the ARGO-YBJ data is -2.61$\pm$0.04.
With this analysis for the first time a ground-based measurement of the CR spectrum overlaps data obtained with direct methods for more than one energy decade.

This measurement has been extended to higher energies exploiting an hybrid measurement with a prototype of the future Wide Field of view Cherenkov Telescope Array (WFCTA) of the LHAASO project \cite{lhaaso}.
The telescope, located at the south-east corner of the ARGO-YBJ detector, about 78.9 m away from the center of the RPC array, is equipped with 16$\times$16 photomultipliers (PMTs), has a FOV of 14$^{\circ}\times$16$^{\circ}$ with a pixel size of approximately 1$^{\circ}$ \cite{nim2011}.

From December 2010 to February 2012, in a total exposure time of 728,000 seconds, the ARGO-YBJ/WFCTA system collected and reconstructed 8218 events above 100 TeV according to the following selection criteria: (1) reconstructed shower core position located well inside the ARGO-YBJ central carpet, excluding an outer region 2 m large; (2) more than 1000 fired pads on the central carpet; (3) more than 6 fired pixels in the PMT matrix; (4) a space angle between the incident direction of the shower and the telescope main axis less than 6$^{\circ}$.
This selection guarantees that the Cherenkov images are fully contained in the FOV, an angular resolution better than 0.2$^{\circ}$ and a shower core position resolution less than 2 m.

%%%%%%%%%%%%%%%%%%%%%%%%%%%%%%%%%%%%%%%%
\begin{figure}
\begin{minipage}[ht]{.47\linewidth}
\vspace{-1cm}
  \centerline{\includegraphics[width=\textwidth]{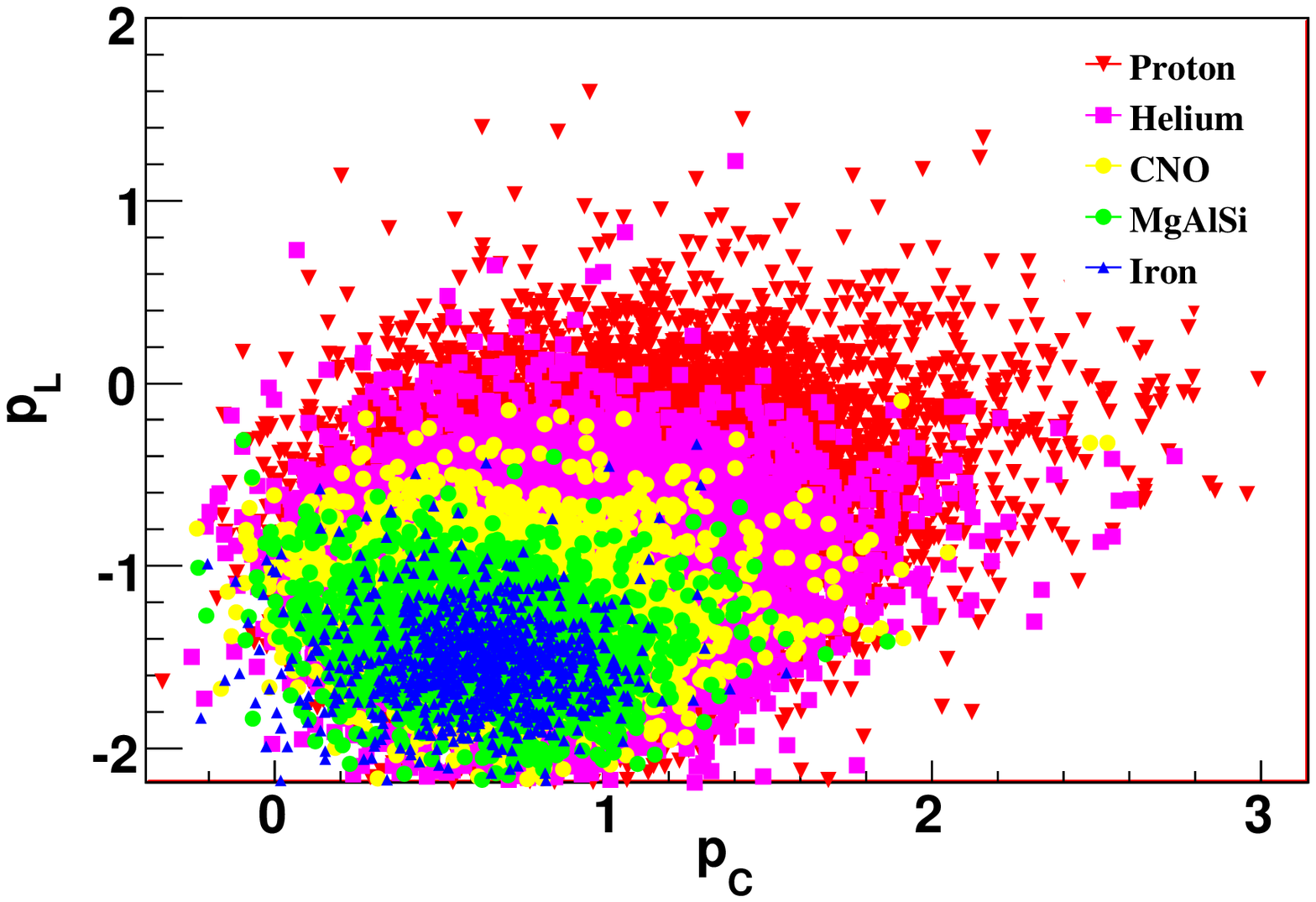}}
    \caption{Scatter plot of the parameters $p_C$ and $p_L$ for showers induced by different nuclei. The primary masses have been simulated in the same fraction, assuming a -2.7 spectral index in the energy range 10 TeV -- 10 PeV.}
\label{fig:pl-pc}
\end{minipage}\hfill
\begin{minipage}[ht]{.47\linewidth}
  \centerline{\includegraphics[width=\textwidth]{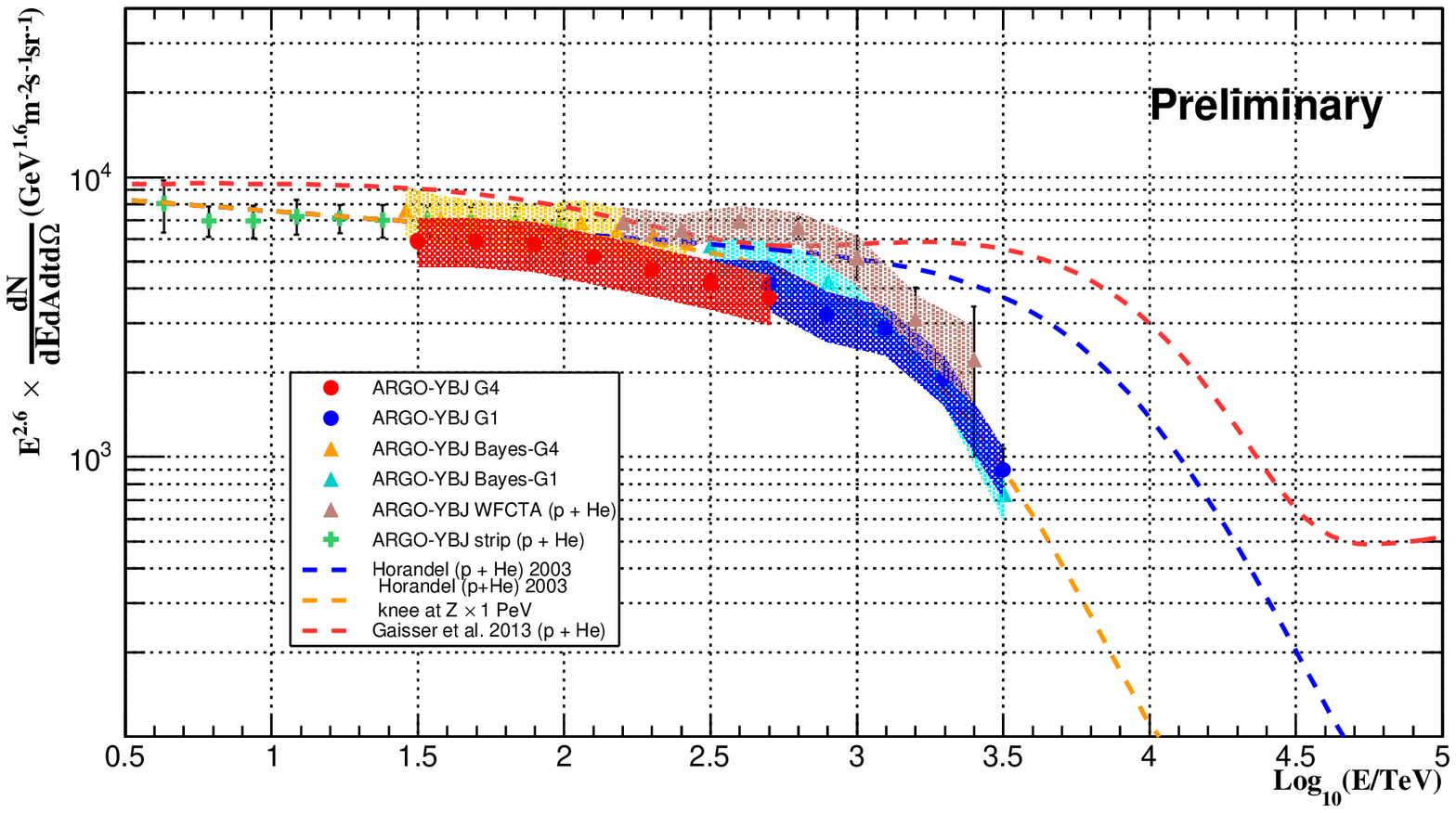} }
    \caption{Light (p+He) component energy spectrum of primary CRs measured by ARGO-YBJ with different analyses. The systematic uncertainty is shown by the shaded area and the statistical one by the error bars. The parametrizations given in \cite{horandel,gst} are shown for comparison. A Horandel-like spectrum with a modified knee at Z$\times$1 PeV is also shown (for details see \cite{discia14}).}
  \label{fig:light}
        \end{minipage}\hfill
\end{figure}
%%%%%%%%%%%%%%%%%%%%%%%%%%%%%%%%%%%%%%%%

According to the MC simulations, the largest number of particles N$_{max}$ recorded by a RPC in a given shower is a useful parameter to measure the particle density in the shower core region, i.e. within 3 m from the core position.
For a given energy, in showers induced by heavy nuclei N$_{max}$ is smaller than in showers induced by light particles. Therefore, N$_{max}$ is a parameter useful to select different primary masses.
In addition, N$_{max}$ is proportional to E$_{rec}^{1.44}$, where E$_{rec}$ is the shower primary energy reconstructed using the Cherenkov telescope.
We can define a new parameter p$_L$ = $log_{10} (N_{max}) - 1.44\cdot log_{10} (E_{rec}/TeV)$ by removing the energy dependence \cite{argo-wfcta}.

The Cherenkov footprint of a shower can be described by the well-known Hillas parameters \cite{hillas85}, i.e. by the width and the length of the image. 
Older showers which develop higher in the atmosphere, such as iron-induced events, have Cherenkov images more stretched, i.e. narrower and longer, with respect to younger events due to light particles which develop deeper.
Therefore, the ratio between the length and the width (L/W) of the Cherenkov image is expected to be another good estimator of the primary elemental composition.

Elongated images can be produced, not only by different nuclei, but also by showers with the core position far away from the telescope, or by energetic showers, due to the elongation of the cascade processes in the atmosphere. 
Simulations show that the ratio of L/W is nearly proportional to the shower impact parameters R$_p$, the distance between the telescope and the core position, which must be accurately measured.
An accurate determination of the shower geometry is crucial for the energy measurement. In fact, the number of photoelectrons collected in the image recorded by the Cherenkov telescope N$_{pe}$ varies dramatically with the impact parameter R$_p$, because of the rapid falling off of the lateral distribution of the Cherenkov light. 
Only an accurate measurement of the shower impact parameters R$_p$, and a good reconstruction of the primary energy allow to disentangle different effects.
A shower core position resolution better than 2 m and an angular resolution better than 0.2$^{\circ}$, due to the high-granularity of the ARGO-YBJ full coverage carpet, allow to reconstruct the shower primary energy with a resolution of 25\%, by using the total number of photoelectrons N$_{pe}$. The uncertainty in absolute energy scale is estimated about 10\%.

Therefore, in order to select the different masses we can define another new parameter p$_C$ = $L/W - 0.0091\cdot(R_p/1\>m) - 0.14\cdot log_{10}(E_{rec} /TeV)$ by removing both the effects due to the shower distance and to the energy.
The values of these parameters for showers induced by different nuclei are shown in the Fig. \ref{fig:pl-pc}. As can be seen from the figure, a suitable selection in the p$_L$ -- p$_C$ space allows to pick out a light composition sample with high purity. In fact, by cutting off the concentrated heavy cluster in the lower-left region in the scatter plot, i.e. p$_L\leqslant$ -0.91 and p$_C \leqslant$ 1.3, the contamination of nuclei heavier than He is less than 5\%.  About 30\% of H and He survives the selection criteria.
In the sample of 8218 events recorded above 100 TeV by the hybrid system, 1392 showers are selected in the (p+He) sub-sample.
The aperture of the ARGO-YBJ/WFCTA system is $\sim$170 m$^2$sr above 100 TeV and shrinks to $\sim$50 m$^2$sr after the selection of the (p+He) component.

The light component energy spectrum measured by the ARGO-YBJ/WFCTA hybrid system is shown in the Fig. \ref{fig:light_spectrum} by the filled red squares. A systematic uncertainty in the absolute flux of 15\% is shown by the shaded area. The error bars show the statistical errors only.
The spectrum can be described by a power law with a spectral index of -2.63 $\pm$ 0.06. The absolute flux at 400 TeV is (1.79$\pm$0.16)$\times$10$^{-11}$ GeV$^{-1}$ m$^{-2}$ sr$^{-1}$ s$^{-1}$. 

This result is consistent for what concern spectral index and absolute flux with the measurements carried out by ARGO-YBJ below 200 TeV and by CREAM. The flux difference is about 10\% and can be explained with a difference in the experiments energy scale at level of 4\%. The measurement of the west-ward displacement of the Moon shadow under the effect of the geomagnetic field, as a function of the event multiplicity, allowed to calibrate the relation shower size - primary energy, thus calibrating the absolute energy scale of the detector at 10\% level below 20 TeV \cite{bartoli11a}.
Above this energy the overposition with CREAM provides a solid anchorage to the absolute energy scale at few percent level.

In order to extend the measurement of the ARGO-YBJ/WFCTA hybrid experiment to the PeV,  the selection cuts in the p$_L$ -- p$_C$ space have been modified as follows: events for which p$_L\leqslant$ -1.25 and p$_C \leqslant$ 1.1 are rejected. 
The aperture is a factor 2.4 larger.
The contamination increases and the purity of the p+He sample below 700 TeV reduces to 93\%. At 1 PeV the contamination is less than 13\% increasing to 44\% around 6 PeV. About 72\% of p+He events survive the selection criteria.

The preliminary measurement of the light component obtained by ARGO-YBJ with different analyses is summarized in the Fig. \ref{fig:light}. As can be seen, all different approaches show evidence of a knee-like structure in the (p+He) spectrum starting from about 650 TeV, well below the all-particle spectrum knee confirmed by ARGO-YBJ at $\sim$3 PeV  \cite{discia14}. With respect to a single power-law with a spectral index --2.63 the deviation is observed at a level of about 6 s.d. .

More sensitive measurements of the energy spectrum and elemental composition in the range from 10$^{12}$ to 10$^{18}$ eV are required to observe the knees of different nuclei and to investigate in detail the end of the spectrum of Galactic CRs and the contradictory results among different experiments.

%%%%%%%%%%%%%%%%%%%%%%%%%%%%%%%%%%%%%
\section {The Fermi Cocoon in the Cygnus region}
%%%%%%%%%%%%%%%%%%%%%%%%%%%%%%%%%%%%%%
\label{sect:cyg-cocoon}

The Cygnus X region is one of the most luminous region of the Northern $\gamma$-ray sky and it is rich in potential CR accelerator sites, e.g. Wolf Rayet stars, OB associations and SNRs. 
The Fermi collaboration detected a complex extended source, in a position consistent with the source ARGO J2031+4157, attributed to the emission of freshly accelerated CRs interacting with gas and radiation, filling a bubble (a ``cocoon'') carved by stellar winds and multiple supernovae shock waves \cite{ackermann11}.

\begin{figure}
\begin{minipage}[ht]{.48\linewidth}
  \centerline{\includegraphics[width=\textwidth]{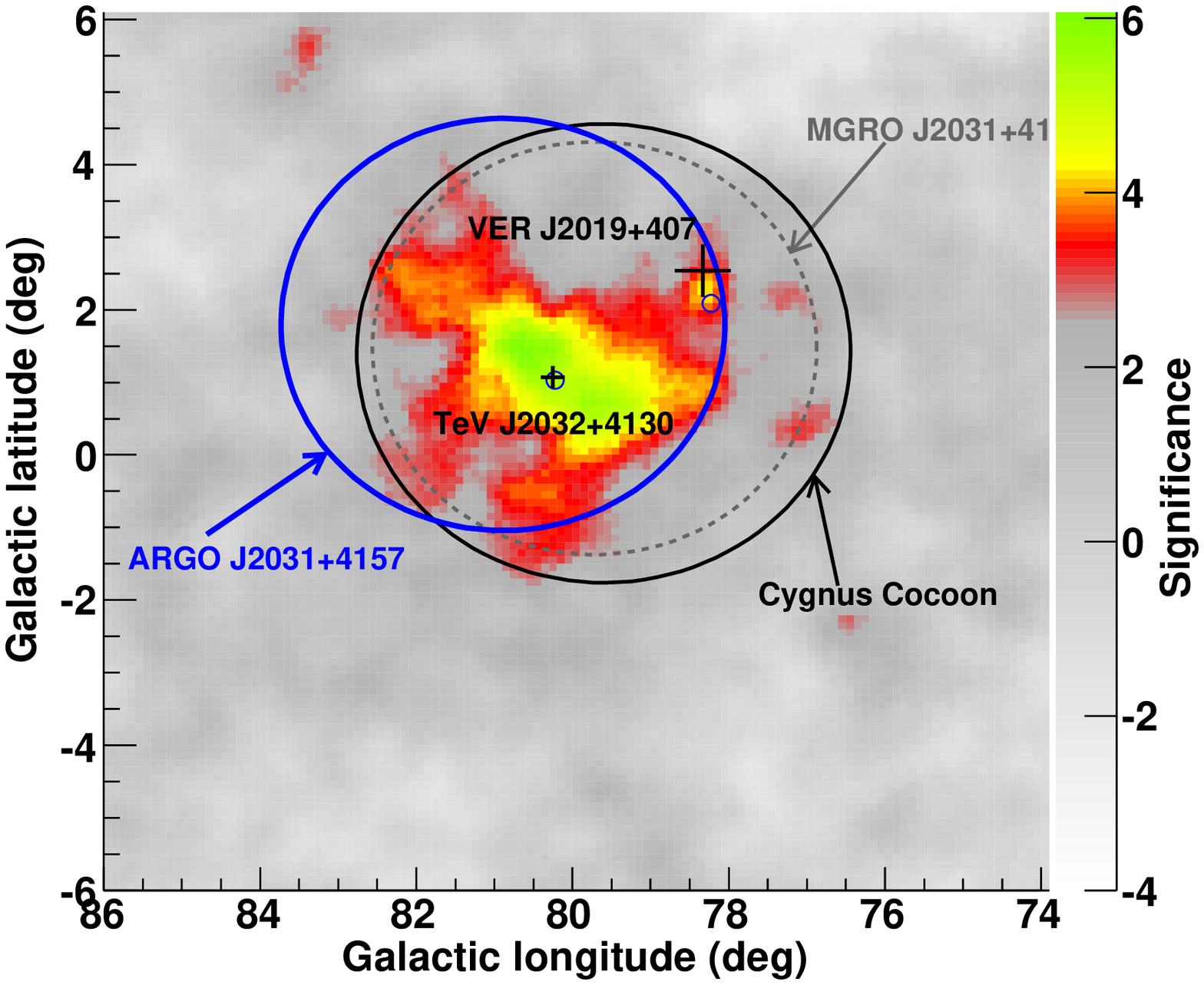}}
\end{minipage}\hfill
\begin{minipage}[ht]{.48\linewidth}
  \centerline{\includegraphics[width=\textwidth]{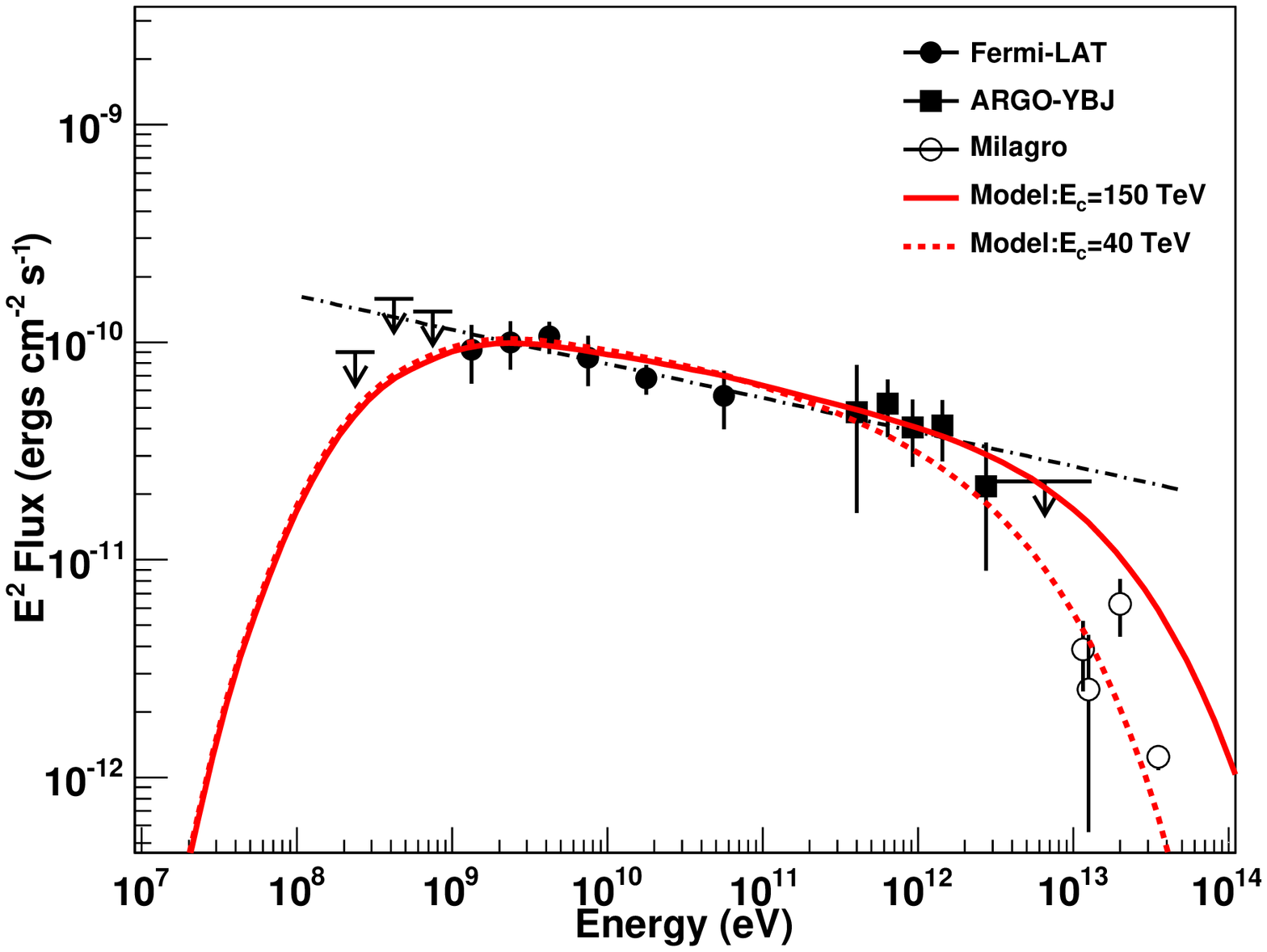} }
        \end{minipage}\hfill
            \caption{Left plot: The significance map of the ARGO J2031+4157 region observed by the ARGO-YBJ experiment. Right plot: Spectral energy distribution of the Cygnus Cocoon. Different markers stand for the spectra measured by different detectors. The dot-dashed line shows the best fit to the Fermi-LAT and ARGO-YBJ data using a simple power-law function. The red solid (dashed) line is obtained by a hadronic model with a proton cutoff energy at 150 (40) TeV. For references and details see \cite{argo-cocoon}.}
  \label{fig:cocoon}
\end{figure}

The significance map around ARGO J2031+4157 as observed by ARGO-YBJ is shown in the left plot of Fig. \ref{fig:cocoon}. 
To improve the sensitivity to $\gamma$-induced events, an optimizated selection based on the shower core position is
applied \cite{argo-cocoon}. A smoothing is applied with an energy-dependent point-spread function evaluated for point $\gamma$-ray sources.
For comparison, the known TeV sources and the Cygnus Cocoon are marked in the figure. The sizes of markers indicate the 68\% containment size of the extension (for details see \cite{argo-cocoon}). 

%%%%%%%%%%%%%%%%%%%%%%%%%%%%%%%%%%%%%%%%
%%%%%%%%%%%%%%%%%%%%%%%%%%%%%%%%%%%%%%%%
%
The energy spectrum of the Cygnus Cocoon measured by Fermi-LAT, ARGO-YBJ and Milagro is shown in the right plot of Fig. \ref{fig:cocoon}. The flux determined by ARGO-YBJ appears consistent with the extrapolation of the Fermi-LAT spectrum suggesting that the emission of ARGO J2031+4157 can be identified as the counterpart of Cygnus Cocoon at TeV energies. 

The combined spectrum of Fermi-LAT and ARGO-YBJ can be described by a differential power law (dot-dashed line) $dN/dE = (3.46\pm 0.33)\times 10^{-9}\cdot (E/0.1\>TeV)^{(-2.16\pm 0.04)}$ photons cm$^{-2}$ s$^{-1}$ TeV$^{-1}$,
suggesting the same origin for both GeV and TeV extended gamma-ray emission. Only statistical errors are shown, the systematic errors on the flux are estimated to be less than 30\%.
The upper limits of Fermi-LAT and ARGO-YBJ indicate the presence of a slope change or cutoff below $\sim$1 GeV and above $\sim$10 TeV, respectively.
At a distance of 1.4 kpc, the observed angular extension of about 2$^{\circ}$ corresponds to more than 50 pc, making the Cygnus Cocoon the largest identified Galactic TeV source. 
The observation of this large TeV source, not removed in the Milagro analysis of the $\gamma$-ray diffuse flux from the Cygnus region, can explain the excess observed by them in the region 65$^{\circ}<$ l $<$85$^{\circ}$, $|b|<$5$^{\circ}$.

Such a large region can be related to different scenarios.
As discussed in \cite{ackermann11}, the favored scenario to explain the emission in the Cygnus Cocoon is the injection of CRs via acceleration from the collective action of multiple shocks from supernovae and the winds of massive stars, which form the Cygnus superbubble. Such superbubbles have been long advocated as CR factories, therefore the Cygnus Cocoon could be the first evidence supporting such hypothesis.

In order to test a possible hadronic origin of the $\gamma$-ray emission through the $\pi^{0}$ decay, we considered inelastic collisions between accelerated protons and target gas. We assumed that the primary protons follow a power law with an index similar to the gamma spectrum and with an exponential cutoff at 150 TeV, as suggested in \cite{ackermann11} to describe CR acceleration by random stellar winds in the Cygnus superbubble. This energy is the maximum  proton cutoff allowed by the ARGO-YBJ upper limit. The resulting spectrum is shown in the right plot of Fig. \ref{fig:cocoon} by the solid red line \cite{argo-cocoon}. 
It is worth noting that the Milagro data are not described by this model and can be reconciled only with a cutoff of about 40 TeV (dashed red line) non consistent with the ARGO-YBJ results.

\section{Conclusions}

The ARGO-YBJ experiment has been in stable data taking for more than 5 years at the YangBaJing Cosmic Ray Observatory. With a duty-cycle greater than 86\% the detector collected about 5$\times$10$^{11}$ events in a wide energy range, from few hundreds GeV up to about 10 PeV, exploiting the performance of the RPCs to image the front of atmospheric showers with unprecedented resolution and detail.  

Since November 2007 to January 2013 ARGO-YBJ monitored the Northern sky at TeV photon energies with a cumulative sensitivity ranging from 0.24 to $\sim$1 Crab units, depending on the declination. This sensitivity, the lowest obtained so far, is allowing a detailed study of different galactic and extragalactic sources, focusing in particular on flaring and extended emissions.

The high granularity of the readout and the high altitude location of the experiment offer a unique opportunity for a detailed study of several characteristics of the hadronic component of the CR flux up to about 10 PeV, starting from an energy window marked by the transition from direct to indirect measurements. In addition, the analog readout of the detector provides a powerful tool to study, for the first time, the particle distribution in the shower core region.
Final analysis with the full statistics of the analog data will give new inputs to the hadronic interaction models currently used to describe particle physics and CRs up to the highest energies.

\end{document}